\documentclass{aastex}
\usepackage{spr-astr-addons}

\begin{document}

\title{Extensive photometry of the intermediate polar MU~Cam: detection of a spin period change}
%% Running heads
\shorttitle{A spin period change in MU~Cam}
\shortauthors{Kozhevnikov}

\author{V. P. Kozhevnikov}
\affil{Astronomical Observatory, Ural Federal University, Lenin Av. 51, Ekaterinburg 
620083, Russia e-mail: valery.kozhevnikov@urfu.ru}

%\email{\emaila}
\begin{abstract} 
Intermediate polars with known rates of spin period changes are not numerous because such tasks require measurements performed for a long time. To measure a spin period change, MU~Cam is a good candidate because it has a spin oscillation with a large amplitude enabling measurements with high precision. Fortunately, in the past the spin period of MU~Cam was measured with high precision. To measure the spin period anew, in 2014--2015 we performed extensive photometric observations of MU~Cam, spanning a total duration of 208~h within 46 nights. We found that the spin, sideband and orbital periods are equal to $1187.16245\pm0.00047$~s, $1276.3424\pm0.0022$~s and $4.71942\pm0.00016$~h, respectively. Comparing the measured spin period with the spin period of MU~Cam in the past, we detected the spin period change with ${\rm d}P/{\rm d}t=-(2.17\pm0.10)\times 10^{-10}$. This rate of the spin period change was not stable and varied in a time scale of years. During four nights in 2014 April--May MU~Cam was fainter than usual by 0.8 mag, and the amplitude of the sideband oscillation was five times larger, denoting significant fraction of disc-overflow accretion. The sideband oscillation showed a double-peaked pulse profile in the normal brightness state. When the star brightness was decreased by 0.8 mag, the sideband oscillation showed a single-peaked pulse profile. In contrast, the spin pulse, which was quasi-sinusoidal, remained remarkably stable both in profile and in amplitude. Moreover, the spin pulse was also remarkably stable in a time scale of years and even decades. MU~Cam is of great interest because it represents a distinctive object with a large and unstable rate of the spin period change and exhibits a distinctive behaviour of the pulse profiles.
\end{abstract} 


\section{Introduction}

Magnetic cataclysmic variables, polars and intermediate polars (IPs), are interacting binary stars, in which accretion occurs onto a magnetic white dwarf. In contrast with polars, intermediate polars (IPs) contain a magnetic white dwarf that spins strongly non-synchronously with the orbital period of the system. Because the magnetic axis does not coincide with the spin axis of the white dwarf, this causes an oscillation with the spin period, which can be observable in optical light, in X-rays and in polarimetry. One more periodic oscillation can appear with the beat period, $1/P_{\rm beat} = 1/P_{\rm spin} - 1/P_{\rm orb}$. The oscillation with the beat period is called the orbital sideband. A natural reason for the sideband oscillation, when it occurs in optical light, is the reprocessing of X-rays at some part of the system that rotates with the orbital period. A review of IPs is given in \citet{Patterson94}.

The usual criteria for IP classification are optical and X-ray spin oscillations. These oscillations must show a high degree of coherence to distinguish them from quasi-periodic oscillations. Especially it is important for optical oscillations because X-ray observations by their nature cannot be very long and because quasi-periodic oscillations in X-rays are less probable. Many theoretical works make assumption that IPs are in spin equilibrium. This, however, is questionable. Therefore long-term tracking of the spin period is an important task because allows an observational test of spin equilibrium due to alternating spin-up and spin-down \citep{Patterson94}. When an IP is not in spin equilibrium, the rate of the spin period change gives understanding of the angular momentum flows within the system \citep{King99}.

MU Cam was identified as an intermediate polar by \citet{Araujo03} and \citet{Staude03}. Optical photometry showed variability with two periods, which were tentatively identified with the orbital period of the system and the spin period of the white dwarf, $P_{\rm orb}= 4.719\pm0.006$~h and $P_{\rm spin}=1187.246\pm0.004$~s, respectively \citep{Staude03}. The latter period was the dominant signal in the hard X-rays. This proves that this period is indeed the spin period of the white dwarf but not the sideband period \citep{Staude08}.

\citet{Yun11} analysed times of maxima of the spin oscillation of MU~Cam, which were presented in \citet{Araujo03}, \citet{Staude03}, \citet{Kim05} and their own times of maxima obtained in 2005--2006.  The (O--C) diagram revealed the significant variation of the spin period of MU~Cam. Although, from this diagram, the behaviour of the spin period in the future was unclear, \citeauthor{Yun11} reported ${\rm d}P/{\rm d}t \approx -4.08 \times 10^{-8}$ (see their page 11). This extremely large ${\rm d}P/{\rm d}t$ seems entirely impossible because an oscillation with such ${\rm d}P/{\rm d}t$ must reveal a low degree of coherence.  None the less, the (O--C) diagram presented in Yun et al. suggests a measurable change of the spin period. Therefore we can attempt to find the spin period change by using high-precision measurements of the spin period obtained in different moments, which are separated by large time intervals. Fortunately, in 2002 \citet{Staude03} measured the spin period with high precision. To measure the spin period in 2014--2015, we performed extensive photometric observations of MU~Cam. In addition, these observations allowed us to find out interesting details of the behaviour of the spin pulse profile and of the sideband pulse profile. In this paper we present results of our extensive photometric observations, which have a total duration of 208~h and cover 15 months.

\section{Observations} \label{observations}

\begin{table}[t!]
%\small
\footnotesize
\caption{Journal of the observations.}
\label{journal}
\begin{tabular}{@{}l c c}
\tableline
\noalign{\smallskip}
Date (UT)   &  BJD$_{\rm UTC}$ start (--2450000) & Length (h)  \\
\noalign{\smallskip}
\tableline

\noalign{\smallskip}
2014 Feb. 22   & 6711.327604 & 3.3 \\
2014 Feb. 23   & 6712.164357 & 4.6 \\
2014 Mar. 1   & 6718.152123  & 9.0 \\
2014 Mar. 2   & 6719.148853 & 2.1 \\
2014 Mar. 3   & 6720.198778 & 3.9 \\
2014 Mar. 4   & 6721.158120 & 8.3 \\
2014 Mar. 21     & 6738.179675 & 5.5 \\
2014 Mar. 23    & 6740.178699 & 7.7 \\
2014 Mar. 24    & 6741.173325 & 6.3 \\
2014 Mar. 25    & 6742.180494 & 3.8 \\
2014 Mar. 26   & 6743.339145 & 4.0 \\
2014 Mar. 27    & 6744.389740 & 2.7 \\
2014 Apr. 1    & 6749.189763 & 6.5 \\
2014 Apr. 3    & 6751.191986 & 4.3 \\
2014 Apr. 18    & 6766.219247 & 4.5 \\
2014 Apr. 19    & 6767.225016 & 4.6 \\
2014 Apr. 21   & 6769.226927 & 4.8 \\
2014 Apr. 22    & 6770.228148 & 3.1 \\
2014 Apr. 26    & 6774.323583 & 1.8 \\
2014 Apr. 30    & 6778.251015 & 1.8 \\
2014 May 2    & 6780.255849 & 3.7 \\
2014 May 4    & 6782.263216 & 3.3 \\
2014 May 5   & 6783.274651 & 2.9 \\
2015 Feb. 13   & 7067.302195 & 6.3 \\
2015 Feb. 14   & 7068.204794 & 3.6 \\
2015 Feb. 16   & 7070.197795  & 8.4 \\
2015 Feb. 17   & 7071.120983 & 3.9 \\
2015 Feb. 18   & 7072.125110 & 10.2 \\
2015 Feb. 19   & 7073.120522 & 2.6  \\
2015 Feb. 20     & 7074.148826 & 1.8 \\
2015 Feb. 25    & 7079.248468 & 7.6 \\
2015 Mar. 10    & 7092.188435 & 3.5 \\
2015 Mar. 11    & 7093.150438 & 3.3 \\
2015 Mar. 13   & 7095.158729 & 5.3 \\
2015 Mar. 14    & 7096.160269 & 8.4 \\
2015 Mar. 16    & 7098.160094 & 8.6 \\
2015 Mar. 17    & 7099.272101 & 5.6 \\
2015 Mar. 18    & 7100.232552 & 6.8 \\
2015 Mar. 26    & 7108.190183 & 7.5 \\
2015 Apr. 15   & 7128.221829 & 1.4 \\
2015 Apr. 19    & 7132.220454 & 1.7 \\
2015 Apr. 22    & 7135.242679 & 1.5 \\
2015 Apr. 23    & 7136.341738 & 1.5 \\
2015 Apr. 25   & 7138.341206 & 2.1 \\
2015 May 10    & 7153.287896 & 2.1 \\
2015 May 11   & 7154.283716 & 1.7 \\
\noalign{\smallskip}
\tableline
\end{tabular}
\end{table}

In observations of variable stars we use a multi-channel pulse-counting photometer with photomultipliers, which makes it possible to perform brightness measurements of two stars and the sky background simultaneously. The design of the photometer is described in \citet{kozhevnikoviz}. Advantages of this photometer in observations of IPs were demonstrated in our previous works where we either discovered optical oscillations for the first time \citep{Kozhevnikov01} or obtained more precise oscillation periods \citep{Kozhevnikov12, Kozhevnikov14}. In these works one can find details of our methods of observations and analysis. Here, we note that, as usual in our observations of IPs, data of MU~Cam were obtained in white light (approximately 300--800~nm). The time resolution was somewhat higher and equal to 4 s. This time resolution seems excessive for MU~Cam, but such an excessive time resolution allows us to fill gaps in observations more accurately and diminishes the errors of the oscillation periods.

\begin{figure}[t!]
\includegraphics[width=84mm]{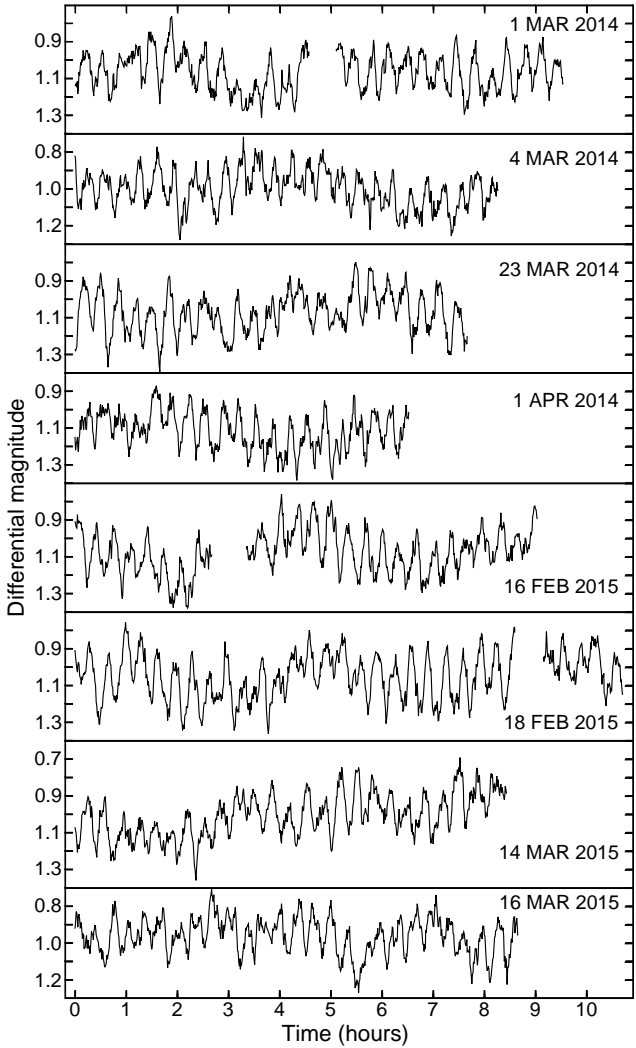}
\caption{Longest differential light curves of MU Cam.}
\label{figure1}
\end{figure}

\begin{figure}[t!]
\includegraphics[width=84mm]{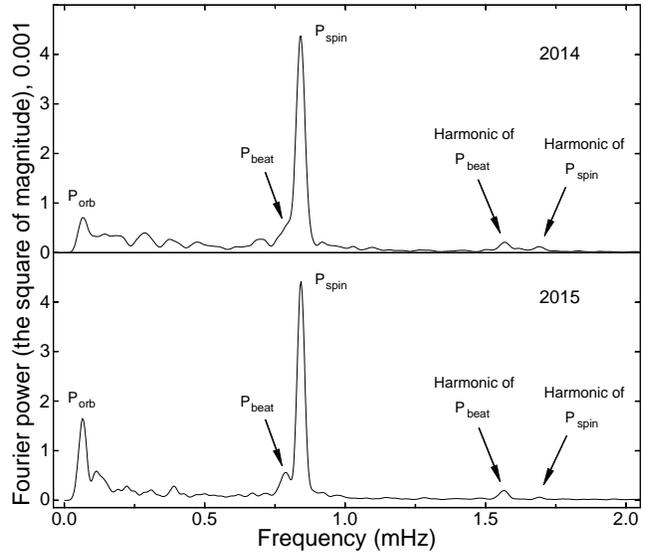}
\caption{Average power spectra calculated by the weighted averaging of 12 power spectra of longest individual light curves of 2014 and of 11 power spectra of longest individual light curves of 2015 from MU~Cam.}
\label{figure2}
\end{figure}

The photometric observations of MU~Cam  were performed in 2014 February--May over 23 nights and in 2015 February--May also over 23 nights with the 70-cm telescope at Kourovka Observatory, Ural Federal University. A journal of the observations is given in Table~\ref{journal}. Note that this table contains BJD$_{\rm UTC}$ which is the  Barycentric Julian Date in the most commonly-used time standard, the Coordinated Universal Time (UTC). Because, in all previous observations of MU~Cam, HJD or BJD are quoted in an unspecified time standard, we believe that this standard is also UTC. Then, our BJD$_{\rm UTC}$ are compatible with those HJD or BJD within an accuracy of a few seconds because of  the high ecliptic latitude  of MU Cam ($+50\degr$) and because only three leap seconds were applied between the first observations of MU Cam and our latest observations. If necessary, one can easily convert our BJD$_{\rm UTC}$ into BJD$_{\rm TDB}$, the Barycentric Julian Date in the Barycentric Dynamical Time standard, by adding 67 s \citep[e.g.,][]{Eastman10}.

The used comparison star is USNO-A2.0 1575-02540930. It has $B=13.7$~mag and $B-R=+0.1$~mag. The colour index of this star is close to the colour index of MU~Cam, $B-R=-0.2$~mag. This minimizes the effect of differential extinction. The noise of the obtained differential light curves is dominated by the photon noise because MU Cam is a faint star of about 15~mag. Fig.~\ref{figure1} presents the longest differential light curves of MU~Cam, in which magnitudes averaged over 40-s time intervals.  The photon noise of these light curves (rms) is in the range 0.013--0.017~mag.

\begin{figure}[t!]
\includegraphics[width=84mm]{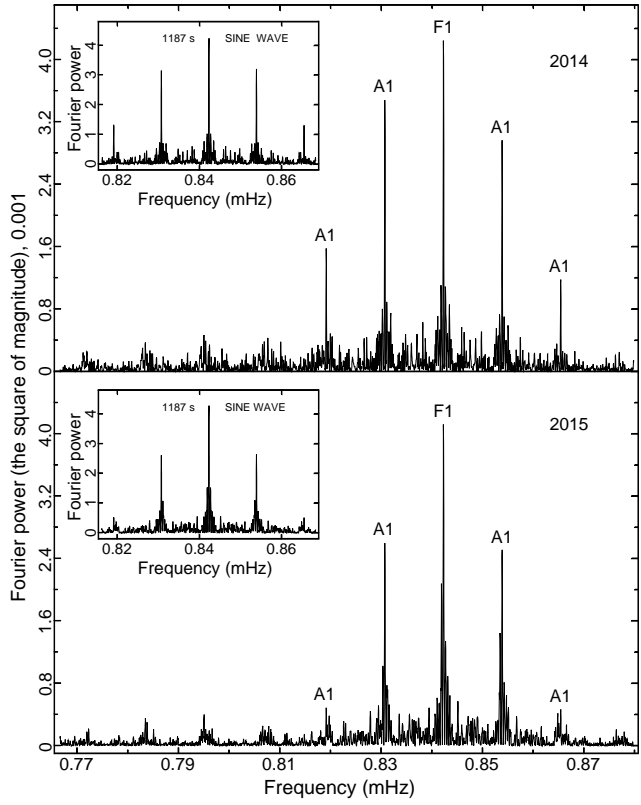}
\caption{Power spectra calculated for the data of 2014 and 2015 from MU~Cam. Inserted frames show the window functions. The principal peaks and one-day aliases of the spin oscillation are labelled with 'F1' and 'A1', respectively.}
\label{figure3}
\end{figure}

\begin{figure}[t!]
\includegraphics[width=84mm]{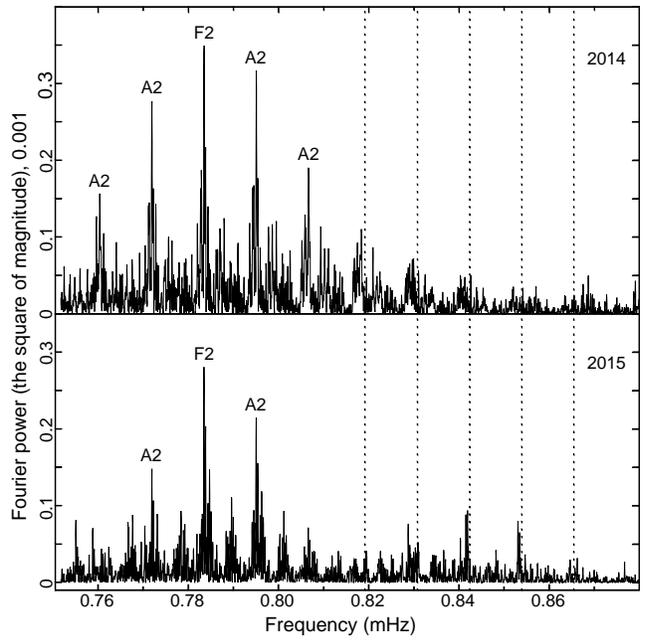}
\caption{Power spectra of the data of MU~Cam, from which the spin oscillation was subtracted. The dotted lines mark the location of the principal peak of the subtracted oscillation and its one-day aliases. The Principal peak and one-day aliases of the sideband oscillation are labelled with 'F2' and 'A2', respectively.}
\label{figure4}
\end{figure}

\section{Analysis and results}

In addition to flickering, which is typical of all types of cataclysmic variables, the light curves of MU~Cam (Fig.~\ref{figure1}) show a periodic oscillation. Obviously, this oscillation having large amplitude corresponds to $P_{\rm spin}$. Fourier analysis allows us to detect a few other periodic oscillations.  Fig.~\ref{figure2} presents the average power spectra calculated by the weighted averaging of 12 power spectra of longest individual light curves of 2014 and of 11 power spectra of longest individual light curves of 2015 from MU~Cam. In addition to the oscillation with $P_{\rm spin}$, these power spectra show oscillations with $P_{\rm beat}$ and $P_{\rm orb}$. High-frequency harmonics of $P_{\rm spin}$ and $P_{\rm beat}$ are also detectable. Note that the harmonic of $P_{\rm beat}$ is strong. This means that the pulse profile of the sideband oscillation is non-sinusoidal.

\begin{figure}[t!]
\includegraphics[width=84mm]{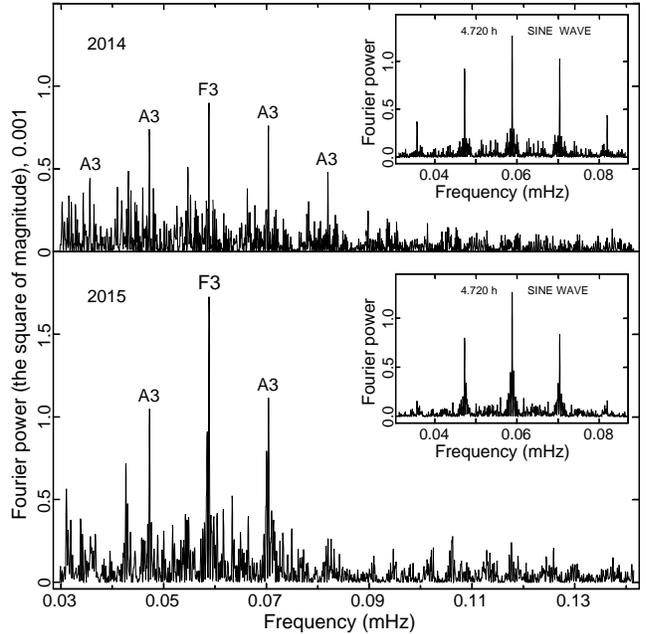}
\caption{Low-frequency parts of the power spectra of the data of MU~Cam, which reveals the peaks corresponding to the orbital period (labelled with 'F3').  The one-day aliases are also present (labelled with 'A3').}
\label{figure5}
\end{figure}

Although the average power spectra presented in Fig.~\ref{figure2} give the overview of the content of the periodic oscillations  seen in MU~Cam and of their relative amplitudes, these spectra are not suitable to find precise periods because their frequency resolution is very low.  Therefore we analysed data incorporated into common time series, the power spectra of which have much better frequency resolution. Fig.~\ref{figure3} presents the Fourier power spectra of two common time series consisting of the data of 2014 and of the data of 2015 in the vicinity of $P_{\rm spin}$. As seen, the spin oscillation reveals the principal peaks and one-day aliases according to the window functions shown in the insets. Small peaks visible between the principal peaks and one-day aliases track the fine structure of the window functions. This strengthens the impression that the spin oscillation is entirely coherent during sufficiently long time spans. From these power spectra, applying a Gaussian function fit to upper parts of the principal peaks, we found that the spin periods are equal to $1187.1673\pm0.0050$ and $1187.1688\pm0.0045$~s in 2014 and 2015, respectively. The errors of the spin periods are found by the method of \citet{schwarzenberg91}.  In our previous works, we made sure that the errors found according to \citeauthor{schwarzenberg91} are true rms errors \citep{Kozhevnikov12, Kozhevnikov14}. Obviously, these periods are compatible with each other. The oscillation semi-amplitudes found from the power spectra are equal to 93 and 91 mmag in 2014 and 2015, respectively.

The power spectra (Fig.~\ref{figure3}) show additional small peaks in the vicinity of the spin oscillation (on the left).  This suggests the presence of the sideband oscillation. These additional peaks, however, do not conform to the window function because they are clustered in groups consisting of many peaks of similar height. Obviously, the sideband oscillation is affected by the spin oscillation, which has close frequency and much higher amplitude. To eliminate the influence of the spin oscillation, we subtracted the spin oscillation from the data. This is facilitated by the fact that the spin oscillation has much higher amplitude and, therefore, is not affected by the sideband oscillation. The obtained power spectra (Fig.~\ref{figure4}) allowed us to detect the sideband oscillation due to the presence of the one-day aliases, which are distributed in frequency and in height as the window functions suggest. The sideband periods are equal to $1276.287\pm0.025$ and $1276.310\pm0.019$~s in 2014 and 2015, respectively. These periods also conform to each other. The semi-amplitudes of the sideband oscillation, which are found from the power spectra, are equal to 27 and 24 mmag in 2014 and 2015, respectively.

To analyse the high-frequency oscillations, we eliminated the low-frequency trends from the light curves by subtraction of a first- or second-order polynomial fit. This is a standard procedure in Fourier analysis and does not affect most frequencies. However, such subtraction diminishes the orbital variability because many our light curves are shorter than the long orbital period of MU~Cam. Therefore, to analyse the orbital variability, we subtracted only nightly averages from the individual light curves.  The corresponding power spectra are presented in Fig.~\ref{figure5}. Although they show significant noise level due to discontinuity and trends in the individual light curves, none the less, these power spectra reveal the principal peaks and one-day aliases, which are caused by the orbital variability. From these power spectra, we found that the orbital periods are equal to $4.7215\pm0.0018$ and $4.7194\pm0.0018$~h in 2014 and 2015, respectively. These periods are also consistent with each other. The semi-amplitudes of the orbital variability, which are found from the power spectra, are equal to 42 and 59 mmag in 2014 and 2015, respectively.

\begin{figure}[t!] 
\includegraphics[width=84mm]{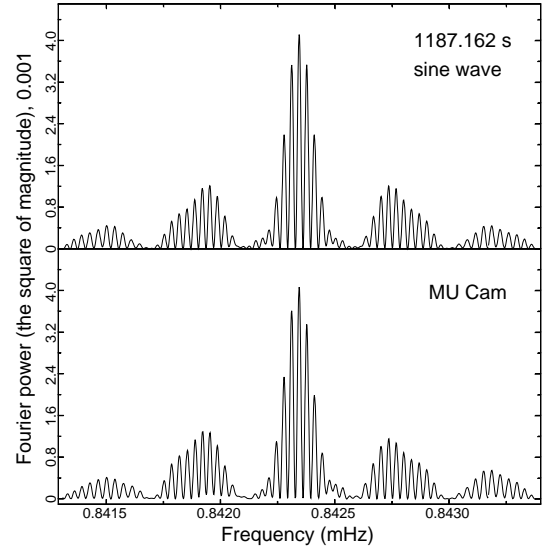}
\caption{Segment of the power spectrum calculated for all data from MU~Cam in the vicinity of $P_{\rm spin}$. The upper frame shows the window function.}

\label{figure6}
\end{figure}

Obviously, more precise oscillation periods can be obtained from all data incorporated into common time series. In such a case, however, identification of principal peaks may be difficult due to complexity of the window function. Although the power spectrum, which is calculated for all data from MU~Cam, reveals a quite simple window function, the difference of the heights of the principal peak and nearest alias is small (Fig.~\ref{figure6}). None the less, in the case of the spin oscillation, the signal-to-noise ratio (S/N) is very high and equal roughly 1000, and, therefore, the aliasing problem is absent.  This is proved by the comparison of the spin period derived from all data, $P_{\rm spin}=1187.16245\pm0.00047$~s, and the spin periods obtained from the data 2014 and from the data of 2015 taken separately. Indeed, the deviations of the periods are less than $1.4\sigma$ when we consider the largest peak as the principal peak. However, if we consider the nearest alias as the principal peak, then the deviations turn out 8--9$\sigma$. This proves the absence of the aliasing problem in the power spectrum of all data of MU~Cam in the case of $P_{\rm spin}$. 

In the vicinity of $P_{\rm beat}$, the power spectrum of all pre-whitened data incorporated into the common time series is similar to the power spectrum in the vicinity of $P_{\rm spin}$. In this case, however, it is difficult to identify the principal peak, because, due to low oscillation amplitude, the S/N is much worse and equal to 90. The periods of two largest peaks are $1276.2857\pm0.0019$ and $1276.3424\pm0.0022$~s. Their semi-amplitudes are nearly equal to each other (25 and 24 mmag). In contrast to $P_{\rm spin}$, both periods are consistent with the sideband periods obtained from the data of 2014 and from the data of 2015 taken separately because the deviations are less than 2.2$\sigma$.
\begin{table}[t!]
\small
\caption{Periods of the detected oscillations.}
%\footnotesize
\scriptsize
\label{table2}
\begin{tabular}{@{}l c c c c }
\tableline
\noalign{\smallskip}
Oscillat. & 2014 & 2015 & 2014+2015 \\
\noalign{\smallskip}
\tableline
\noalign{\smallskip}
Orbital & 4.7215(18) h & 4.7194(18) h & 4.71942(16) h \\ 
Beat & 1276.287(25) s & 1276.310(19) s & 1276.3424(22) s \\
Spin & 1187.1673(50) s & 1187.1688(45) s & 1187.16245(47) s \\
\noalign{\smallskip}
\tableline
\end{tabular}
\end{table}

Although the amplitude of the orbital variability is larger than the amplitude of the sideband oscillation, in the case of $P_{\rm orb}$, the power spectrum of all data also reveal an aliasing problem due to a low S/N of 35. Indeed, in the vicinity of $P_{\rm orb}$, two largest peaks in the power spectrum of all data have periods of $4.71942\pm 0.00016$~h and $4.72212\pm 0.00018$~h and nearly equal semi-amplitudes of 50 and 49 mmag. Both periods conform to the orbital periods obtained from the data of 2014 and from the data of 2015 taken separately because the deviations are less than 1.5$\sigma$. Fortunately, \citet{Kim05} measured the orbital period of MU~Cam with high precision, $P_{\rm orb}=4.71943\pm0.00008$~h. One of our periods nearly strictly coincides with this period whereas another period differs by $14\sigma$. Thus, the true orbital period, which follows from all our data, is equal to $4.71942\pm 0.00016$~h.  Using the relation $1/P_{\rm beat} = 1/P_{\rm spin} - 1/P_{\rm orb}$, we found that the true sideband period is equal to $1276.3424\pm0.0022$~s. Table~\ref{table2} presents all periods, which we found in our data of MU~Cam.

The high precision of $P_{\rm spin}$ allows us to derive an oscillation ephemeris with a long validity. However, a rather large noise level in the individual light curves makes it impossible to find oscillation phases directly. Then, we might obtain oscillation phases from folded light curves. However, phases of the oscillation with $P_{\rm spin}$ can be shifted due to interaction with the sideband oscillation when light curves are short compared to $P_{\rm orb}$ \citep[e.g.,][]{Warner86}. Therefore, we found the times of spin maxima from the folding of large portions of data, in which the effect of the sideband oscillation is negligible.  We obtained the initial time from all data and utilized the data subdivided into four groups (see Table~\ref{table3}) for verification. To find the times of maxima, we used a Gaussian function fitted to upper parts of the maxima seen in the folded light curves. Next, we placed these times in the middle of corresponding observations. Finally, for the oscillation with $P_{\rm spin}$, we obtained the following ephemeris:
{\small
\begin{equation}
{\rm BJD(max)}=2456932.839815(22)+0.0137403061(54) {\it E}.
\label{ephemeris}
\end{equation} }

\begin{table}[t!]
%\small
\scriptsize
\caption{Verification of the spin ephemeris.}
\label{table3}
\begin{tabular}{@{}l c c c c}
\tableline
\noalign{\smallskip}
Time     & BJD(max)             & N. of    & O--C $ \times 10^{3}$     \\
interval     & (-245\,5000)      & cycles  & (days)    \\
\noalign{\smallskip}
\tableline
\noalign{\smallskip}
22 Feb--26 Mar 2014    & 6727.422255(36)    & --14950     &  +0.02(9)   \\ 
27 Mar--5 May 2014      & 6763.903042(88)    & --12295     &  +0.29(11)   \\
13 Feb--13 Mar 2015    & 7081.344856(27)     & +10808     &  --0.19(7)     \\
14 Mar--11 May 2015    & 7125.259160(100)   & +14004    &  +0.10(13)   \\
\noalign{\smallskip}
\tableline
\end{tabular}
\end{table}

\begin{figure}[hptb]
\includegraphics[width=84mm]{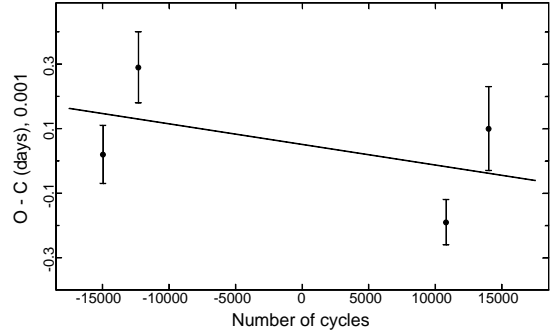}
\caption{(O--C) diagram for all data from MU~Cam, which are subdivided into four groups and folded with $P_{\rm spin}$. Although the O--C diagram shows an appreciable slope, this slope is not significant.}
\label{figure7}
\end{figure}

\begin{figure*}[t!]
\includegraphics[width=174mm]{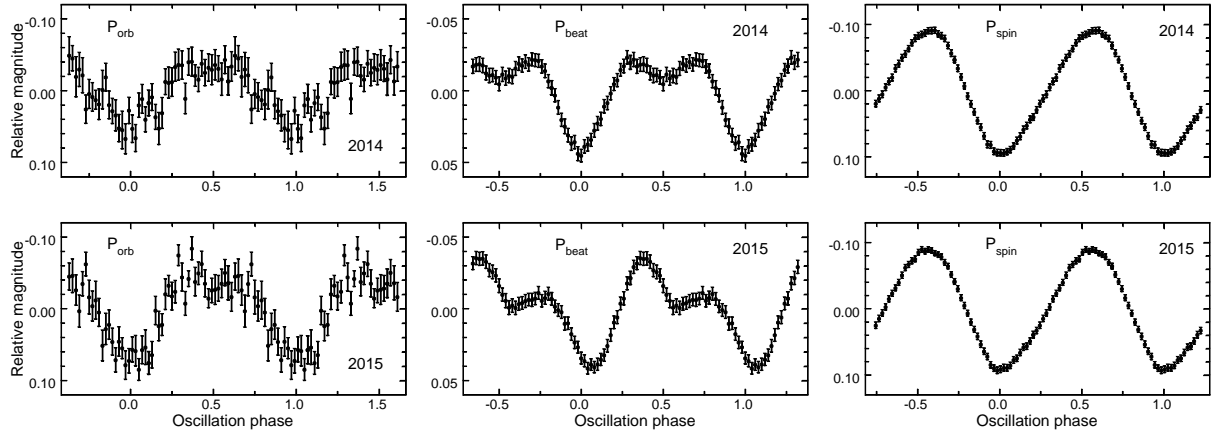}
\caption{Pulse profiles of three oscillations obtained for the data of 2014 and 2015 from MU~Cam. The oscillation with $P_{\rm orb}$ (on the left) shows a distinct minimum. This minimum, however, cannot be caused by an eclipse, because it is too wide. The oscillation with $P_{\rm beat}$ (in the middle) reveals a double-peaked pulse profile, which is variable in a time scale of years.  The oscillation with $P_{\rm spin}$ (on the right) demonstrates a remarkably stable asymmetric pulse profile with a slow rise and a rapid decline.}
\label{figure8}
\end{figure*}

According to this ephemeris, we derived the (O--C) values and numbers of the oscillation cycles for the four portions of data and presented them in Table~\ref{table3}. The (O--C) values (Fig.~\ref{figure7}) obey the relation: ${\rm (O-C)}=0.000051(110) - 0.0000000064(81) {\it E}$. Using this relation, we might correct the slope and displacement along the vertical axis, which are visible in Fig.~\ref{figure7}. However, because all parameters in this relation are less than their rms errors, such a correction seems unnecessary.  Therefore, we leave the ephemeris unchanged. According to the rms error of the spin period, the formal validity of this ephemeris is equal to 90 years (a $1\sigma$ confidence level).

Fig.~\ref{figure8} presents the pulse profiles of the oscillations observed in MU~Cam. The orbital light curve (on the left) shows a distinct minimum. This minimum, however, cannot be caused by an eclipse, because it is too wide \citep[e.g.,][]{Patterson98b}. The sideband oscillation (in the middle) reveals a double-peaked pulse profile, which is variable in a time scale of years.  Because this profile is non-sinusoidal, the sideband oscillation produces a strong first harmonic (Fig.~\ref{figure2}). The spin oscillation (on the right) demonstrates a remarkably stable asymmetric pulse profile with a slow rise and a rapid decline. Because this profile is quasi-sinusoidal, the spin oscillation reveals only a weak first harmonic (Fig.~\ref{figure2}).

\section{Discussion}

Analysing (O--C) values for the linear spin ephemeris given by \citet{Staude03}, \citet{Yun11} detected the spin period change of MU~Cam. Although their new cubic ephemeris showed a trend to spin-up, the behaviour of the spin period in the future was unclear. Moreover, by using a cubic ephemeris, it seems inappropriate to derive ${\rm d}P/{\rm d}t$ because a cubic term of an ephemeris means that ${\rm d}P/{\rm d}t$ varies. However, for MU~Cam, Yun et al. reported ${\rm d}P/{\rm d}t \approx -4.08 \times 10^{-8}$ (see their page 11). This rate of the spin period change perplexes because it is roughly 100 times greater than the maximum rate of spin period changes observed in other IPs \citep[see table~1 in][]{Warner96}. In addition, with such a large ${\rm d}P/{\rm d}t$, an oscillation must reveal a low degree of coherence, which is not typical of periodic oscillations seen in IPs.

\subsection{Spin period change}

To detect changes of the spin period of MU~Cam directly from measurements of $P_{\rm spin}$, we performed extensive photometric observations during 46 nights in 2014--2015. From the analysis of all these data, we obtained $P_{\rm spin} = 1187.16245\pm0.00047$~s. Comparing this period with the spin period found by \citet{Staude03} 12 years ago, which was equal to $1187.246\pm0.004$~s, we derived ${\rm d}P/{\rm d}t=-(2.17\pm0.10)\times 10^{-10}$. This detection of the spin period change possesses a high level of confidence because ${\rm d}P/{\rm d}t$ is 22 times greater than its rms error.

In 2004 March--May we performed photometric observations of MU~Cam with the same technique \citep{Kozhevnikov06}. Now we used those data to confirm the spin period change, which we detected by using the results of  \citeauthor{Staude03} Unfortunately, 11 years ago we had insufficient experience and evaluated the spin period only roughly with the random error equal to the  half-width of the peak at half-maximum. Obviously, this error is a conservative error. Now we reanalysed those data using the methods described in the present work and found $P_{\rm spin} = 1187.240\pm0.010$~s. However, as seen in Fig.~6 in \citet{Kozhevnikov06}, in 2004 March--May the sideband oscillation showed much higher amplitude and, therefore, noticeably affected  the spin oscillation.  From numerical experiments with artificial time series we found that, due to the effect of the sideband oscillation, the rms error of $P_{\rm spin}$ must be increased roughly by 30 per cent. Thus, in 2004 March--May the spin period of MU~Cam was equal to $1187.240\pm0.013$~s.  Using this period and the spin period obtained in the present work, we found ${\rm d}P/{\rm d}t=-(2.29\pm0.39) \times 10^{-10}$. Although, in this case, the rms error is somewhat larger, the confidence level of this detection of the spin period change is also high (6$\sigma$). Moreover, this ${\rm d}P/{\rm d}t$ agrees with the ${\rm d}P/{\rm d}t$ obtained by using the data of \citet{Staude03}.

The precision of the spin periods from the data of 2014 and from the data of 2015 taken separately is sufficient to detect the spin period change in MU~Cam. Then, we can consider two independent pairs of observations. The first pair is the data of \citeauthor{Staude03} and our data of 2014. The second pair is our data of 2004 and our data of 2015. For these pairs of data, we obtained ${\rm d}P/{\rm d}t=-(2.13\pm0.17) \times 10^{-10}$ and ${\rm d} P/{\rm d}t=-(2.06\pm0.41) \times 10^{-10}$, respectively. These values of ${\rm d}P/{\rm d}t$ agree with each other, where their confidence levels are also sufficiently high (12$\sigma$ and $5\sigma$).  Thus, we can say that our detection of the spin period change is confirmed by independent observations.

\citet{Staude03},  \citet{Kim05} and \citet{Yun11} obtained times of maxima of the spin oscillation in 2002--2006. In addition, from our folded light curves of 2004, we found $BJD(\rm max)=2453110.86947(7)$.  Unfortunately, using our ephemeris and these times of maxima, we cannot check the evaluated ${\rm d}P/{\rm d}t$ because the (O--C) values exceed many oscillation cycles. We, however, can analyse only the data of 2002--2006 using the following formula \citep{Breger98}:
\begin{equation}
{\rm (O-C)}=0.5 \, \frac{1}{p} \, \frac{{\rm d}P}{{\rm d}t} \, t^2.
\label{breger}
\end{equation}
Using formula \ref{breger}, we calculated an artificial (O--C) diagram with the initial time of the ephemeris and with the spin period in \citet{Staude03} and with our ${\rm d}P/{\rm d}t$. Next, we compared this artificial diagram with the (O--C) diagram, which was calculated according to the linear ephemeris in \citet{Staude03} and was shown in Fig.~2 in \citet{Yun11}. For the first cluster of points, which presents the data of \citeauthor{Staude03} and \cite{Araujo03}, the (O--C) values do not differ systematically because the time intervals are too short. For the second and third clusters of points, which present the data of \citet{Kim05}, the average $\Delta$(O--C) are equal to --0.13 cycles and --0.10 cycles, accordingly. Here, $\Delta$(O--C) is the (O--C) value according to the ephemeris in \citeauthor{Staude03} minus the artificial (O--C) value. Negative signs of $\Delta$(O--C) mean that during 2002--2005 the rate of the spin period change was somewhat higher than it follows from our ${\rm d}P/{\rm d}t$.

For our data of 2004, which are located close to the second cluster of points, we obtained $\Delta$(O--C)=+0.001 cycles. Obviously, this nearly strict coincidence of the (O--C) values is accidental. Indeed, in the case of our observations of 2004, the systematic error caused by the sideband oscillation \citep[e.g.,][]{Warner86}  is appreciable and equal to $\pm0.015$ cycles. Therefore, this coincidence can be only accidental. None the less, the difference between the $\Delta$(O--C) for the data of \citeauthor{Kim05} and the $\Delta$(O--C) for our data of 2004 is significant and exceeds 0.1 cycles.  One of the reasons might be the effect of different methods used to find times of maxima. \citeauthor{Kim05} used a sine wave fit, but we used a Gaussian function fit. However, in the case of the asymmetric spin pulse profile of MU~Cam,  the difference of times of maxima caused by these different methods, is less than 0.04 cycles and cannot account for the contradiction between the data of  \citeauthor{Kim05}  and our data of 2004. Moreover, this difference is difficult to account for by changes in the geometry of the accretion flow \citep[e.g.,][]{Patterson98a} because such effects are observable in individual light curves, but we consider average data of several light curves.

For the fourth cluster of points, which presents the data of \citeauthor{Yun11}, the average $\Delta$(O--C) is equal to +0.18 cycles. Then, we inevitably come to the conclusion that the rate of the spin period change was decreased between 2005 and 2006. Thus, the (O--C) diagram presented in Fig.~2 in \citeauthor{Yun11} roughly conforms to ${\rm d}P/{\rm d}t=-(2.17\pm0.10)\times 10^{-10}$, but the ${\rm d}P/{\rm d}t$ is not stable and varies in a time scale of years. However, ${\rm d}P/{\rm d}t \approx -4.08 \times 10^{-8}$ reported by \citeauthor{Yun11} entirely contradicts the  diagram presented in their Fig.~2. Hence, this extremely large ${\rm d}P/{\rm d}t$ should be considered erroneous.

The analysis of our data confirms the conclusion made by \citeauthor{Yun11} that MU~Cam reveals the spin period change. However, taking into account that the ${\rm d}P/{\rm d}t$ reported by \citeauthor{Yun11} is erroneous, our detection of the spin period change of MU~Cam should be considered as a new result. Although our ${\rm d}P/{\rm d}t=-(2.17\pm0.10)\times 10^{-10}$ is  large, it is comparable with the ${\rm d}P/{\rm d}t$, which were observed in other IPs. Whereas many IPs show ${\rm d}P/{\rm d}t$ of an order of $10^{-11} - 10^{-12}$, a few IPs reveal ${\rm d}P/{\rm d}t$ of an order of $10^{-10}$ \citep[see, e.g., table~1 in][]{Warner96}. These are FO~Aqr \citep[${\rm d}P/{\rm d}t=-2.0\times10^{-10}$,][]{Williams03} and PQ~Gem \citep[${\rm d}P/{\rm d}t=+1.1\times10^{-10}$,][]{Mason97}. In addition, recently we detected the spin period change of V647~Aur, where ${\rm d}P/{\rm d}t=-1.36\times10^{-10}$ \citep{Kozhevnikov14}. From table ~1 in \citet{Warner96}, we can conclude that IPs with known ${\rm d}P/{\rm d}t$ are sparse. Hence, our detection of the spin period change in MU~Cam is an important result.

Our observations together with previous observations reveal that, on the average, the spin period of MU~Cam is decreasing. The average rate of the spin period change of MU~Cam should be measured by future observations. Often, such researches are performed by using an ephemeris and (O--C) values. However, in the case of MU~Cam, such a method seems not quite suitable because the rate of the spin period change varies. In addition, the large rate of the spin period change makes it impossible to use a linear ephemeris within large time intervals.  Indeed, our ephemeris \ref{ephemeris} have a formal validity of 90 years. But, in fact, this validity is reduced up to 3.4 years due to the large rate of the spin period change. To measure the rate of the spin period change by another way, one can use direct measurements of $P_{\rm spin}$. This is facilitated by the large amplitude of the spin oscillation. Indeed,  our results show that photometric observations, which are obtained within roughly 100 hours and cover a few months, are sufficient to achieve the precision required for detection of changes of the spin period of MU~Cam in a time scale of decades.

\subsection{Behaviour of the oscillation profiles}
In contrast with polars, in which accretion occurs through a stream onto one pole of the white dwarf, IPs normally exhibit accretion through a truncated accretion disc. Therefore two opposite poles of the white dwarf accretes simultaneously. This, however, does not mean that the pulse profile of the spin oscillation must be necessarily double-peaked even if both poles of the white dwarf can be visible to the observer. \citet{Norton99} argued that in IPs with strong magnetic fields two accreting poles act in phase. Then, the spin pulse profile must be always single-peaked. On the contrary, IPs with weak magnetic fields have two accreting poles, which act in anti-phase. Therefore, depending on the geometry, such IPs can produce both double-peaked and single-peaked spin pulse profiles. \citeauthor{Norton99} also argued that IPs with $P_{\rm spin} <700$~s have weak magnetic fields and therefore often reveal double-peaked spin pulse profiles. According to $P_{\rm spin}$, MU~Cam must have a strong magnetic field. Therefore it is not surprising that this IP show a single-peaked spin pulse profile (Fig.~\ref{figure7} on the right). We, however, must note that, in spite of a short spin period of 465~s, V455~And reveals a strikingly similar single-peaked spin pulse profile. Both profiles are asymmetric with a slow rise and with a rapid decline \citep[compare Fig.~5 in][]{Kozhevnikov12}  and Fig.~\ref{figure7} in the present work).

As seen in Fig.~\ref{figure7} (on the right), the spin pulse is very stable both in profile and in amplitude between 2014 and 2015. Moreover, 11 years ago the spin pulse had exactly the same asymmetric profile with the same amplitude \citep[see Fig.~6 in][]{Kozhevnikov06}. Such an asymmetry can arise if the white dwarf is not in spin equilibrium because material flowing from the disc can be mainly behind or ahead of the magnetic axis depending on the character of deviation from the spin equilibrium \citep{Evans04, Vrielmann05, Evans06}. MU~Cam does not show alternating spin-up and spin-down and, therefore, is not in spin equilibrium \citep[e.g.,][]{Williams03}. None the less, variations of ${\rm d}P/{\rm d}t$ mean that the accretion  flow and torque fluctuate. Moreover, the significant brightness variations of MU~Cam in a time scale of years, which were noted by \citet{Kim05} and \citet{Staude08}, also mean that the accretion flow fluctuates. The very high stability of the pulse profile of the spin oscillation in large time scales is difficult to explain under the conditions when the accretion flow and torque acting upon the white dwarf are unstable and therefore must cause changes of the form of the accretion regions emitting optical light.

\begin{figure}[t!]
\includegraphics[width=84mm]{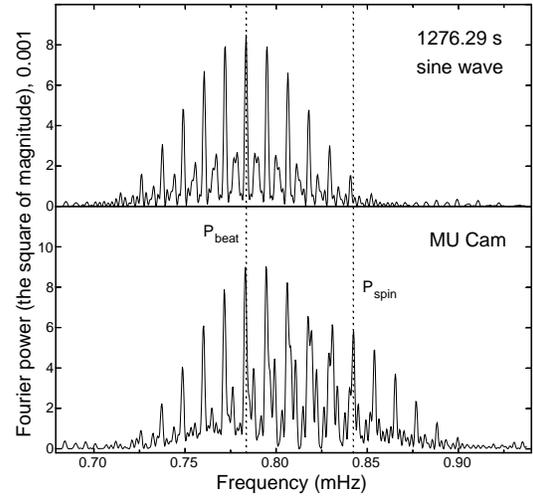}
\caption{Power spectrum of the data obtained in 2014 April 30--May 5 when the brightness  of MU~Cam was decreased by 0.8 mag. The amplitude of the sideband oscillation was greatly increased whereas the amplitude of the spin oscillation revealed only a little change. The upper frame shows the window function. The dotted lines mark $P_{\rm beat}$ and $P_{\rm spin}$.}
\label{figure9}
\end{figure}

\begin{figure}[t!]
\includegraphics[width=84mm]{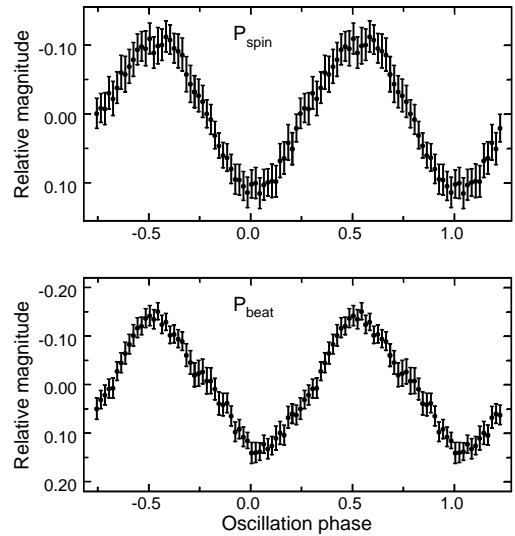}
\caption{Pulse profiles of the spin oscillation (on the top) and sideband oscillation (on the bottom), which were obtained from the data of 2014 April 30--May 5 when MU~Cam decreased its brightness by 0.8 mag. The pulse profile of the spin oscillation remained the same as usual, whereas the pulse profile of the sideband oscillation, instead of double-peaked, became single-peaked and revealed much higher amplitude.}
\label{figure10}
\end{figure}

The well-established canonical interpretation of the optical orbital sideband consists in the reprocessing of X-rays at some structure of the system that rotates with the orbital period. This structure can be the secondary star or hot spot, which arises in the place where the accreting material impacts the disc. The reprocessing, however, cannot account for a strong orbital sideband in X-rays. The X-ray orbital sideband can be accounted for by alternation of the accretion flow between two poles of the white dwarf with the sideband frequency when an IP, in addition to disc-fed accretion, demonstrates disc-overflow accretion \citep{Wynn92}. MU~Cam represents such a rare exemplar. \citet{Staude08} observed the change of the accretion mode in MU~Cam. In the normal brightness state, this IP revealed the disc-fed accretion. When the star brightness was reduced by a magnitude, in addition to the disc-fed accretion, the disc-overflow accretion appeared. This was proved by the appearance of the orbital sideband in X-rays. In addition, in the optical light, the amplitude of the sideband oscillation was greatly increased.

Although, as seen in Fig.~\ref{figure1}, during  our observations  MU~Cam was bright and had $B \approx 14.8$~mag, we found a quite long episode when MU~Cam was noticeably fainter than usual and had $B \approx 15.6$~mag. This episode covers the four latter observing nights of 2014. The power spectrum obtained from these data is presented in Fig.~\ref{figure9}. As seen, the amplitude of the sideband oscillation increased by five times, whereas the amplitude of the spin oscillation remained nearly the same. The change of the star brightness and the corresponding behaviour of the sideband oscillation resemble the results described in \citeauthor{Staude08} and hence denote significant fraction of disc-overflow accretion during the four latter nights of 2014.

Fig.~\ref{figure10} (on the bottom) shows the pulse profile of the sideband oscillation, which was obtained from the four latter nights of 2014. As seen, the pulse profile is single-peaked. Moreover, the phase of the pulse maximum (0.5) coincide with the phase of the local minimum of the pulse profile obtained from the data of 2014 and 2015 (Fig.~\ref{figure7} in the middle). Although the length of the four latter light curves of 2014 is much less than the length of all light curves of 2014, this pulse maximum having large height must influence onto the average pulse profile. Indeed, excluding these four light curves from the data of 2014, we made sure that the pulse profile changed. The local minimum between two maxima became much deeper, and the pulse profile of the sideband oscillation in 2014 became similar to the pulse profile of the sideband oscillation in 2015, although the heights of the two local maxima differed considerably.  Thus, the interplay between disc-fed accretion and disc-overflow accretion can change the pulse profile of the sideband oscillation. 

However, in the case of low star brightness, the pulse maximum of the sideband oscillation (phase 0.5) does not coincide with any of two maxima, which this pulse profile shows in the case of high star brightness (phases 0.25 and 0.75). Therefore, it seems unlikely that the double-peaked pulse profile of the sideband oscillation is produced by superposition of two oscillations, which are caused by the reprocessing of X-rays and by disc-overflow accretion. Hence, this double-peaked pulse profile can be caused by the reprocessing of X-rays either at a complicated structure of the disc or at the secondary star and the hot spot of the disc.

The pulse profile and the amplitude of the sideband oscillation strongly change when the star brightness changes.  In contrast, the pulse of the spin oscillation remains stable both in profile and in amplitude (compare Fig.~\ref{figure7} on the right and Fig.~\ref{figure10} on the top). This behaviour seems instructive.  The change of the star brightness by 0.8 mag means that the light intensity emitted by the disc was changed by two times. Because the amplitude and pulse profile of the spin oscillation remained practically unchanged, the structure of the accretion regions emitting optical light could not change, but the light intensity emitted by them must change also by two times. It seems surprising that the accretion regions emitting optical light do not change their structure when their light intensity changes by two times. Moreover, it seems also surprising that the light intensity of the disc and the light intensity of these regions change in equal proportion.   

\section{Conclusions}
We performed extensive photometric observations of  MU~Cam over 46 nights in 2014 and 2015 and obtained the following results:
\begin{enumerate}
\item Due to the large observational coverage and low relative noise level, we evaluated the spin period of the white dwarf with high precision. The spin period is equal to $1187.16245\pm0.00047$~s. The semi-amplitude of the spin oscillation is 90~mmag.
\item Comparing this spin period and the spin period, which was found by \citet{Staude03} 12 years ago, we discovered that the spin period of MU~Cam decreases with ${\rm d}P/{\rm d}t=-(2.17\pm0.10)\times 10^{-10}$. The rate of the spin period change is not stable and fluctuates in a time scale of years. 
\item The pulse profile of the spin oscillation is asymmetric with a slow rise and a rapid decline. This pulse is remarkably stable both in profile and in amplitude in time scales of years and even decades. 
\item The profile and amplitude of the spin pulse remained practically unchanged when MU~Cam temporarily decreased its brightness by 0.8 mag. This means that the accretion regions emitting optical light do not change their structure when their light intensity is changed by two times. It also means that the light intensity of the disc and the light intensity of these regions change in equal proportion.
\item In addition to the spin oscillation, we detected the sideband oscillation with a period of $1276.3424\pm0.0022$~s and with an average semi-amplitude of 25~mmag.
\item Generally, the amplitude of the sideband oscillation was 3.5 times less than the amplitude of the spin oscillation. During the four latter observing nights of 2014 the brightness of MU~Cam was decreased by 0.8 mag, and the amplitude of the sideband oscillation was increased by five times, denoting that during these four nights we observed MU~Cam in the state of significant fraction of disc-overflow accretion.  
\item During the normal brightness of MU~Cam, the sideband oscillation showed a double-peaked pulse profile. During the low brightness of MU~Cam, which denotes significant fraction of disc-overflow accretion, the sideband oscillation showed a single-peaked pulse profile.
\end{enumerate}

\section*{Acknowledgments}
This work was supported by Act 211 Government of the Russian Federation, agreement N\textsuperscript{\underline{o}} 02.A03.21.0006. This research has made use of the SIMBAD database, operated at CDS, Strasbourg, France. This research also made use of the NASA Astrophysics Data System (ADS). Author thanks Astrophysics and Space Science for publication of this article.  The final publication is available at Springer via http://dx.doi.org/10.1007/s10509-016-2859-0.

{}

\vspace{1.0cm}
Fig. 1.  Longest differential light curves of MU Cam.

\vspace{0.4cm}
Fig. 2.  Average power spectra calculated by the weighted averaging of 12 power spectra of longest individual light curves of 2014 and of 11 power spectra of longest individual light curves of 2015 from MU~Cam.

\vspace{0.4cm}
Fig. 3.  Power spectra calculated for the data of 2014 and 2015 from MU~Cam. Inserted frames show the window functions. The principal peaks and one-day aliases of the spin oscillation are labelled with 'F1' and 'A1', respectively.

\vspace{0.4cm}
Fig. 4.  Power spectra of the data of MU~Cam, from which the spin oscillation was subtracted. The dotted lines mark the location of the principal peak of the subtracted oscillation and its one-day aliases. The Principal peak and one-day aliases of the sideband oscillation are labelled with 'F2' and 'A2', respectively.

\vspace{0.4cm}
Fig. 5.  Low-frequency parts of the power spectra of the data of MU~Cam, which reveals the peaks corresponding to the orbital period (labelled with 'F3').  The one-day aliases are also present (labelled with 'A3').

\vspace{0.4cm}
Fig. 6.  Segment of the power spectrum calculated for all data from MU~Cam in the vicinity of $P_{\rm spin}$. The upper frame shows the window function.

\vspace{0.4cm}
Fig. 7.  (O--C) diagram for all data from MU~Cam, which are subdivided into four groups and folded with $P_{\rm spin}$. Although the O--C diagram shows an appreciable slope, this slope is not significant.

\vspace{0.4cm}
Fig. 8.  Pulse profiles of three oscillations obtained for the data of 2014 and 2015 from MU~Cam. The oscillation with $P_{\rm orb}$ (on the left) shows a distinct minimum. This minimum, however, cannot be caused by an eclipse, because it is too wide. The oscillation with $P_{\rm beat}$ (in the middle) reveals a double-peaked pulse profile, which is variable in a time scale of years.  The oscillation with $P_{\rm spin}$ (on the right) demonstrates a remarkably stable asymmetric pulse profile with a slow rise and a rapid decline. 

\vspace{0.4cm}
Fig. 9.  Power spectrum of the data obtained in 2014 April 30--May 5 when the brightness of MU~Cam was decreased by 0.8 mag. The amplitude of the sideband oscillation was greatly increased whereas the amplitude of the spin oscillation revealed only a little change. The upper frame shows the window function. The dotted lines mark $P_{\rm beat}$ and $P_{\rm spin}$.

\vspace{0.4cm}
Fig. 10.  Pulse profiles of the spin oscillation (on the top) and sideband oscillation (on the bottom), which were obtained from the data of 2014 April 30--May 5 when MU~Cam decreased its brightness by 0.8 mag. The pulse profile of the spin oscillation remained the same as usual, whereas the pulse profile of the sideband oscillation, instead of double-peaked, became single-peaked and revealed much higher amplitude.
\end{document}